\documentclass[a4paper,12pt,aps,prl,superscriptaddress,reprint,floatfix,showpacs,showkeys]{revtex4-1}

\usepackage[utf8]{inputenc}
\usepackage[english]{babel}
\usepackage[T1]{fontenc}
\usepackage{amsmath,amssymb}
\usepackage{graphicx}
\usepackage{amsfonts}
\usepackage{lmodern}
\usepackage{textcomp}
\usepackage[squaren,Gray]{SIunits}
\usepackage{natbib}
\usepackage{hyperref}
\usepackage{fancyhdr}

\graphicspath{{Images_resoumission/}}

\begin{document} 

\pagestyle{fancy}
\lhead{Published article available online at \url{http://journals.aps.org/prl/abstract/10.1103/PhysRevLett.112.223902}}

\title{Superfilamentation in air}

\author{Guillaume Point}
\email{guillaume.point@ensta-paristech.fr}
\affiliation{Laboratoire d'Optique Appliquée - ENSTA ParisTech, Ecole Polytechnique, CNRS - 828 boulevard des Maréchaux, 91762 Palaiseau Cedex, France}
\author{Yohann Brelet}
\affiliation{Laboratoire d'Optique Appliquée - ENSTA ParisTech, Ecole Polytechnique, CNRS - 828 boulevard des Maréchaux, 91762 Palaiseau Cedex, France}
\author{Aurélien Houard}
\affiliation{Laboratoire d'Optique Appliquée - ENSTA ParisTech, Ecole Polytechnique, CNRS - 828 boulevard des Maréchaux, 91762 Palaiseau Cedex, France}
\author{Vytautas Jukna}
\affiliation{Centre de Physique Théorique - Ecole Polytechnique, CNRS - 1 route de Saclay, 91128 Palaiseau Cedex, France}
\author{Carles Mili{\'a}n}
\affiliation{Centre de Physique Théorique - Ecole Polytechnique, CNRS - 1 route de Saclay, 91128 Palaiseau Cedex, France}
\author{Jérôme Carbonnel}
\affiliation{Laboratoire d'Optique Appliquée - ENSTA ParisTech, Ecole Polytechnique, CNRS - 828 boulevard des Maréchaux, 91762 Palaiseau Cedex, France}
\author{Yi Liu}
\affiliation{Laboratoire d'Optique Appliquée - ENSTA ParisTech, Ecole Polytechnique, CNRS - 828 boulevard des Maréchaux, 91762 Palaiseau Cedex, France}
\author{Arnaud Couairon}
\affiliation{Centre de Physique Théorique - Ecole Polytechnique, CNRS - 1 route de Saclay, 91128 Palaiseau Cedex, France}
\author{André Mysyrowicz}
\email{andre.mysyrowicz@ensta-paristech.fr}
\affiliation{Laboratoire d'Optique Appliquée - ENSTA ParisTech, Ecole Polytechnique, CNRS - 828 boulevard des Maréchaux, 91762 Palaiseau Cedex, France}

\begin{abstract}
The interaction between a large number of laser filaments brought together using weak external focusing leads to the emergence of few filamentary structures reminiscent of standard filaments, but carrying a higher intensity. The resulting plasma is measured to be one order of magnitude denser than for short-scale filaments. This new propagation regime is dubbed \textit{superfilamentation}. Numerical simulations of a nonlinear envelope equation provide good agreement with experiments.\\
\\
Published article: \url{http://journals.aps.org/prl/abstract/10.1103/PhysRevLett.112.223902}
\end{abstract}

\pacs{42.65.Jx, 42.65.Sf, 52.70.Kz}

\maketitle

\textit{Introduction}.--- For the past 20 years, much attention has been given to the study of propagation of short intense laser pulses through transparent media. When the peak power of such pulses exceeds a particular critical power $P_{cr}$ ($P_{cr} \equiv \frac{3.72\lambda_{0}^{2}}{8\pi n_{0}n_{2}} \sim \unit{5}{\giga\watt}$ in air at $\lambda_{0} = \unit{800}{\nano\metre}$ with $n_{0}$ and $n_{2}$, respectively, the linear and Kerr optical indices), filamentation sets in. In this case, the Kerr effect prevails and leads to beam self-focusing. The resulting collapse is stopped by the combination of multiphoton absorption by nitrogen and oxygen molecules in air and the defocusing effect induced by the resulting plasma. A dynamic competition between these opposite trends is then established, and the beam is able to maintain a very high intensity over long distances without significant defocusing \cite{Couairon2007}. When $P > P_{cr}$, total beam collapse occurs, with the generation of a single filament. For $P \gg P_{cr}$, the laser pulse unavoidably breaks down into many filaments. This short-scale filamentation \cite{Centurion2005} is initiated by intensity fluctuations in the beam profile that are amplified by  modulational instability, leading to local self-focusing of sub-beams instead of a total beam collapse \cite{Vidal1996,Mlejnek1999,Fibich2005}.

So far most studies have focused on the characterization of single filaments ($P \gtrsim P_{cr}$) or the interaction between pairs of filaments. Such binary interaction leads to the fusion, repulsion, energy redistribution or spiral motion of the filaments, depending on their relative optical phase \cite{Xi2006,Varma2008,Shim2010}. There is also a large cumulative nonlinear phase shift acquired during these processes so that any initial phase correlation between the filaments is lost \cite{Shim2012}.

In this Letter, we experimentally characterize a dense filament bundle which can no longer be described by binary interactions. This regime is obtained by weakly focusing a terawatt femtosecond laser pulse in air. In the focal region, a reorganization of the interacting filament bundle takes place with the emergence of a few filamentary structures reminiscent of standard filaments, but carrying a significantly higher intensity and persisting over tens of centimeters. The resulting plasma is measured to be one order of magnitude denser than for short-scale filaments. Simulations solving a non-linear envelope equation \cite{Couairon2011} reproduce well these features. This effect is measured to be stable in time with a high shot-to-shot reproducibility. The occurrence of such stable events contrasts with other extreme and unpredictable optical phenomena arising from nonlinear wave-matter interactions, so called \textit{rogue waves}, which have been recently reported \cite{Majus2011,Demircan2012}. This may have impact on the recently discovered laser action in air induced by filaments, where the plasma strings form the active medium \cite{Luo2003,Kartashov2012b,Liu2013a}, but also on the generation of long-lived virtual optical waveguides in the atmosphere \cite{Jhajj2014}.

\begin{figure}[!ht]
\begin{center}
\includegraphics[width = .4 \textwidth]{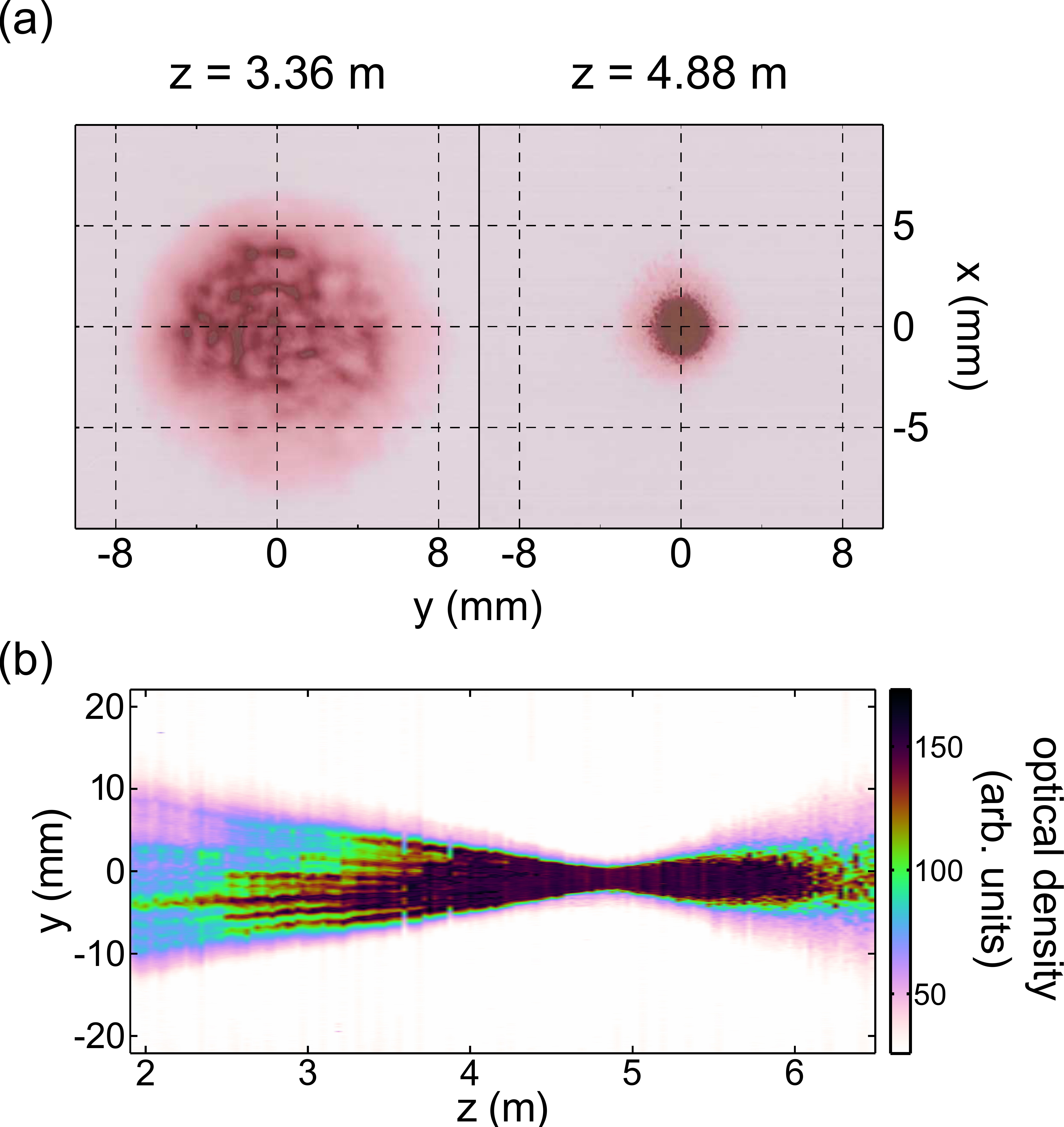}
\end{center}
\caption{(Color online) Experimental investigation of the structure of the multifilament bundle. (\textbf{a}): impacts on photographic paper in the transverse \textsc{xy} plane for $z$ = 3.36, and \unit{4.88}{\metre}. (\textbf{b}): reconstructed side view (\textsc{yz}, $x=0$ plane) of the intensity map using these impacts.}
\label{figure_1}
\end{figure}

\textit{Experimental results}.--- The experimentally studied multifilament bundle was generated by focusing a \unit{200}{\milli\joule}, \unit{30}{\milli\metre} full width at half maximum (FWHM) pulse at \unit{800}{\nano\metre} from the Ti:Sapphire ENSTAmobile laser chain using a \unit{5}{\metre} fused silica converging lens (numerical aperture NA = $3\times10^{-3}$). The pulse duration was adjusted to \unit{175}{\femto\second} in order to avoid white light emission inside the lens. This corresponds to a peak power of about 230 $P_{cr}$. The experiments took place in an air-conditioned room with standard temperature and pressure conditions. In order to record the formation of filaments during the propagation of the laser pulse, we used a method based on the single-shot darkening of photographic plates \cite{Mechain2005b}. Once the propagating laser pulse reaches the intensity required for high-field ionization of air, characteristic circular burns become apparent on the photographic plates (cf. figure \ref{figure_1}-(\textbf{a})), signaling the presence of plasma strings. We recorded such imprints every \unit{2}{\centi\metre} along the beam propagation direction (taken as the \textsc{z} axis, the origin being at the position of the lens), allowing the 3D-reconstruction of the beam undergoing multifilamentation. A longitudinal projection of this reconstruction in the (\textsc{yz}, $x = 0$) plane is given in figure \ref{figure_1}-(\textbf{b}).

The first filaments, displayed in black in the figure, appear around $z = \unit{2.5}{\metre}$. They multiply and converge under the focusing action of the lens, eventually merging together around $z = \unit{4}{\metre}$. After the geometric focus ($z = \unit{5}{\metre}$), they start disappearing with only few of them left at $z = \unit{6.5}{\metre}$. However, this method of detection cannot resolve any structure around the geometrical focus because the photographic paper is completely saturated.

To address this problem, we investigated the luminescence emitted in this compressed plasma region. Side pictures of the plasma luminescence in the focal region were taken using a standard camera (Nikon D50 fitted with a AF-5 Micro NIKKOR \unit{60}{\milli\metre} objective).

\begin{figure}[!ht]
\begin{center}
\includegraphics[width = .4 \textwidth]{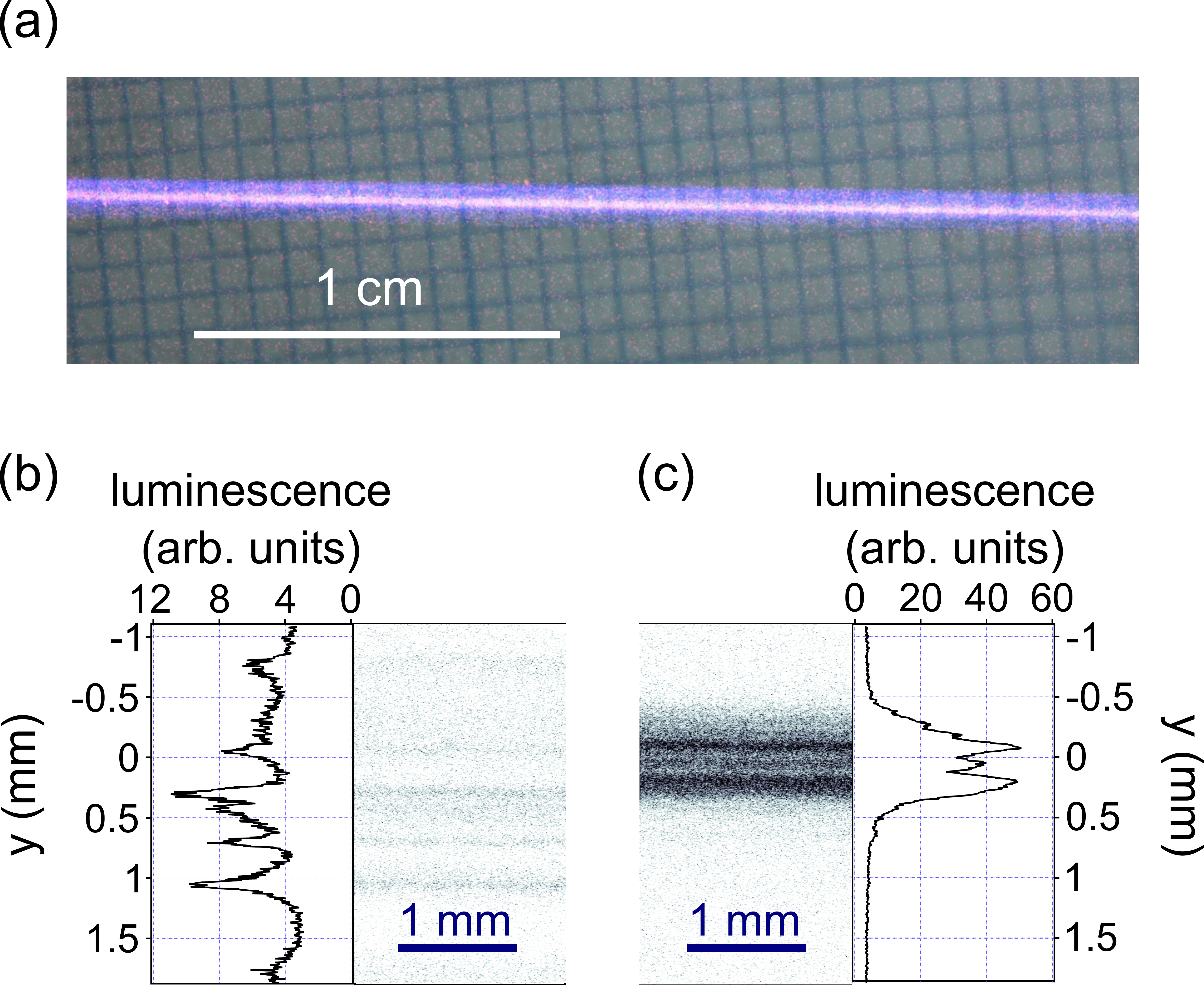}
\end{center}
\caption{(Color online) Luminescence of the plasma channels. (\textbf{a}): side-view picture of the filament bundle in the focal zone. (\textbf{b}): side-view iCCD images of the plasma luminescence centered at $z = \unit{4.40}{\metre}$ and corresponding $z$-integrated luminescence profile. (\textbf{c}): other iCCD picture centered at $z = \unit{4.90}{\metre}$ and corresponding $z$-integrated luminescence profile.}
\label{figure_2}
\end{figure}

As seen in figure \ref{figure_2}-(\textbf{a}), these images reveal a long and thin intense string embedded in a millimeter-wide fluorescent channel. This string maintains a constant diameter over at least several centimeters. To study more quantitatively these intense strings, a time-gated intensified CCD camera (PI-MAX from Princeton Instruments fitted with a 4$\times$ microscope objective and a BG39 colored filter) was used to take single shot, transverse pictures of the multifilament bundle in the focal zone. The spatial resolution of the pictures is \unit{5}{\micro\metre} and the field of view is about \unit{1}{\milli\metre}. The transverse FWHM of these inner plasma structures is the same as that of the single filaments recorded at $z = \unit{4.4}{\metre}$ (cf. figure \ref{figure_2}-(\textbf{b})).

We can go further and use a spectroscopic analysis to characterize the plasma strings in the focal region. The corresponding luminescence spectrum was recorded between 330 and \unit{800}{\nano\metre} using a monochromator (model H-20 UV, Jobin Yvon) and a photomultiplier (model 56 UVP). We averaged all the data over more than 200 laser shots per point. The measured spectrum is similar to the one observed in the case of a single filament, consisting mainly of molecular nitrogen lines, with almost all the signal lying between 330 and \unit{450}{\nano\metre} \cite{Talebpour2001,Xu2009}. No nitrogen or oxygen atomic emission line was detected. It has been shown that, for a single filament, the luminescence signal is directly proportional to the total number of free electrons in the plasma \cite{Xu2009}. We can therefore use the luminescence signal as a diagnostic to evaluate the electron number in the focal region.

\begin{figure}[!ht]
\begin{center}
\includegraphics[width = .4 \textwidth]{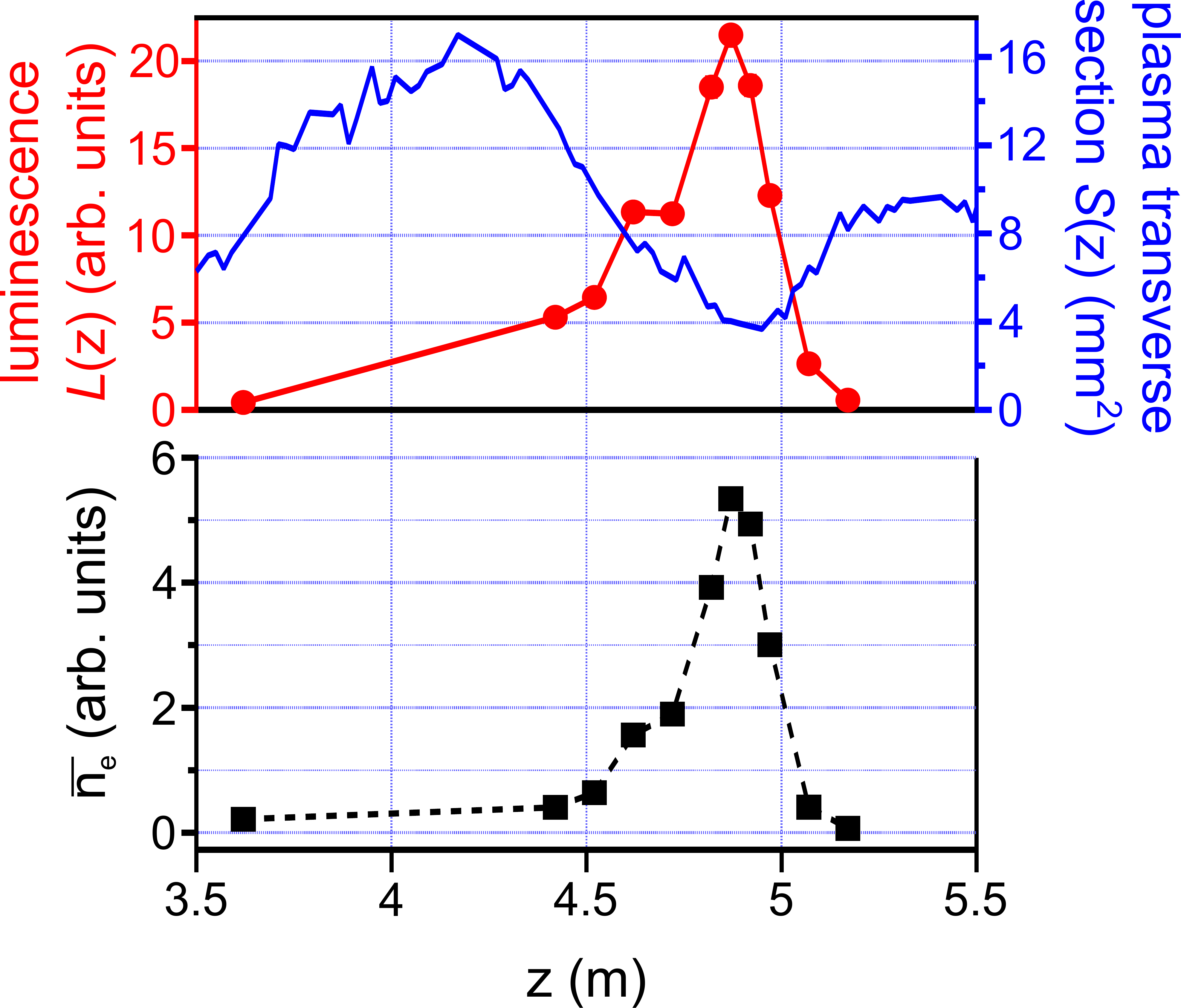}
\end{center}
\caption{(Color online) Evolution of the plasma luminescence (red solid line and circles), the plasma surface (blue solid line) and the average electron density $\overline{n_{e}}$ (black dashed line and squares) with $z$.}
\label{figure_3}
\end{figure}

Figure \ref{figure_3} shows the evolution of the plasma luminescence along $z$ by measuring the cumulative surface of the three most important emission lines of neutral molecular nitrogen, respectively at 3371, 3577 and \unit{3805}{\angstrom} ($C^{3}\Pi_{u}$-$B^{3}\Pi_{g}$ band). It is worth noting the low shot-to-shot fluctuation of the plasma luminescence, with a standard deviation of less than $10 \%$ over more than 200 shots. It indicates that the plasma strings are highly reproducible.

Evaluating the plasma transverse cross-sectional area $S(z)$ along $z$ therefore allows us to estimate the average electron density $\overline{n_{e}}$ in the plasma following this relation:
\begin{equation}
\overline{n_{e}}(z) \propto L(z)/S(z),
\end{equation}
where $L(z)$ is the average recorded plasma luminescence. This is done by correlating the darkening of the previously used photographic plates to the occurrence of ionization \cite{Mechain2005b}. We can define three different regions from this last graph: 
\begin{itemize}
\item the first one, up to $z = \unit{4.5}{\metre}$, is characterized by an almost constant electron density, which can be attributed to the intensity clamping inside short-scale filaments \cite{Kasparian2000,Becker2001}. At the end of this region, filaments are still discernible on the photographic plates, with about 45 being recorded.
\item the second zone sees the quick rise of $\overline{n_{e}}$, reaching a maximum value around $z = \unit{4.9}{\metre}$. This position lies within 5$\%$ of the full-beam collapse distance given by the Marburger formula in the case of a Gaussian pulse \cite{Couairon2007}. The maximum plasma density culminates at about 18 times the plateau level of the previous region.
\item once $z = \unit{5}{\metre}$ is reached, the $L/S$ ratio quickly falls, marking the end of the dense plasma region.
\end{itemize}

This results in a \unit{50}{\centi\metre} long region where the mean plasma density and intensity are significantly higher than the mean values encountered in short-scale filamentation. This length corresponds to about 6 linear Rayleigh lengths, which we measured using the same laser at very low intensity.

This estimation, however, only deals with an average value of the electron density in the plasma. While this method gives insight about the relative evolution of the plasma along $z$, it lacks a reference with which we could estimate the absolute value of the plasma density. On the other hand, the electron density at the core of standard filaments is well known to be on the order of \unit{10^{22}}{\rpcubic\metre} \cite{Chen2010a}. To compare the peak values of the electron density between the dense plasma strings and the short-scale filaments, iCCD pictures displayed in figure \ref{figure_2}-(\textbf{b}) can be used. Figure \ref{figure_3} shows that filaments seen at $z = \unit{4.4}{\metre}$ are still standard clamped filaments, the averaged electron density being at the plateau level characteristic of short-scale filaments. Thus, we have $n_{e,peak}(z = \unit{4.4}{\metre}) \sim \unit{10^{22}}{\rpcubic\metre}$. Because the dense strings seen around $z = \unit{4.9}{\metre}$ have the same FWHM as the short-scale filaments, the ratio of their electron density is directly given by that of their corresponding luminescence levels. This gives a maximum electron density in the focal zone lying between 6 and 11 times the clamping density, that is one order of magnitude higher than this plateau. Thus, the peak density encountered along $z$ is on the order of $\unit{10^{23}}{\rpcubic\metre}$.

\textit{Simulations}.--- We model the time-harmonic propagation of the electric field envelope ${\cal E}(x,y,z)$ in air ($n_0\approx1$) by means of a unidirectional beam propagation equation accounting for diffraction, optical Kerr effect, multiphoton absorption, plasma absorption and defocusing (see, e. g., \cite{Couairon2011} for details):
\begin{equation}
\begin{split}
\frac{\partial {\cal E}}{\partial z} = ~& ic\frac{\partial_x^2+\partial_y^2}{2 \omega_0} {\cal E} + i\frac{\omega_0}{c}n_2 \vert {\cal E}\vert^2{\cal E} \\
&-\left\{\frac{\beta_8}{2} \vert {\cal E}\vert^{14}+\frac{\sigma n_{e}}{2}[1+i\omega_0\tau_c] \right\}{\cal E}.
\end{split}
\label{eq1}
\end{equation}
Here $\omega_0=2\pi c/\lambda_0$ is the carrier frequency, $n_2=\unit{2\times10^{-19}}{\centi\metre\squared\cdot\rp\watt}$, $\beta_8=\unit{8\times10^{-98}}{\power{\centi\metre}{13}\cdot\power{\watt}{-7}}$, $\sigma=\unit{5.6\times10^{-20}}{\centi\metre\squared}$, and $\tau_c\approx\unit{350}{\femto\second}$. Equation \eqref{eq1} was solved numerically by means of a split-step technique \cite{Agrawal2007}. A first step is performed in Fourier ($k_{x},k_{y}$) space to account only for diffraction. A second step, performed in direct ($x,y$) space, accounts for amplitude and phase changes due to nonlinear terms. Because of the absence of temporal dynamics in our modeling, the presence of plasma is accounted for by mapping the electric field intensity to the electron-plasma density generated by optical field ionization and avalanche at the temporal center of a \unit{175}{\femto\second} (pre-chirped) Gaussian pulse with peak intensity $I$. Simulations are initialized with noise levels of $10\%$ in intensity and  $0.2\%$ in phase, mimicking experimental input beam irregularities, triggering modulational instability and the subsequent multifilamentation. The spatial resolution was \unit{6.2}{\micro\metre} in the \textsc{x} and \textsc{y} directions, and \unit{2.5}{\milli\metre} in the \textsc{z} direction, which is sufficient to resolve short scale filamentary structures in a reasonable computational time.

\begin{figure}[!ht]
\begin{center}
\includegraphics[width = .48 \textwidth]{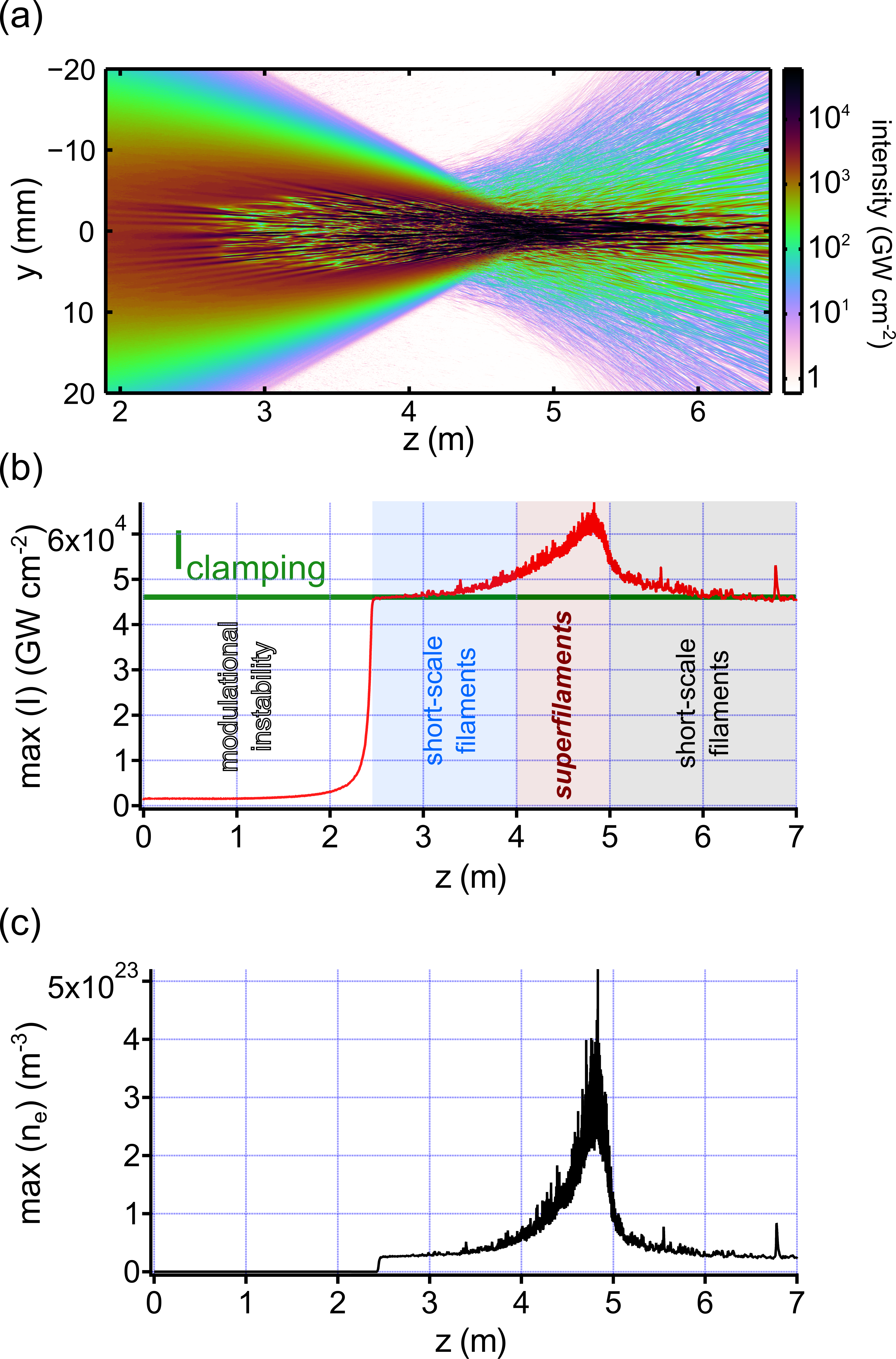}
\end{center}
\caption{(Color online) Simulated evolution of the multifilament bundle. (\textbf{a}): side view (\textsc{yz} plane, $x=0$) of the intensity map. (\textbf{b}): evolution of the maximum intensity in the \textsc{xy} plane with $z$. (\textbf{c}): evolution of the maximum electron density in the \textsc{xy} plane with $z$.}
\label{figure_4}
\end{figure}

The simulation results are presented in figure \ref{figure_4}. Figure \ref{figure_4}-(\textbf{a}) shows the same intensity side-view as in figure \ref{figure_1}-(\textbf{b}). A logarithmic color scale was chosen to mimic the response of the photographic paper. Even without taking a saturation effect into account, simulation and experiment show good agreement with respect to the beam shape, waist at the focus and the observation of only few filaments past $z = \unit{6}{\metre}$.

Figures \ref{figure_4}-(\textbf{b}) and (\textbf{c}) give respectively the simulated maximum intensity and the maximum electron density encountered in the transverse \textsc{xy} plane at a given $z$ position. Both values are very low before the onset of the first filaments, which appear around $z = \unit{2.5}{\metre}$. These short-scale filaments present the typical intensity and plasma density clamping values reported in the literature ($I \sim \unit{4.5\times 10^{13}}{\watt\cdot\centi\metre\rpsquared}, n_{e} \sim \unit{2.5\times10^{22}}{\rpcubic\metre}$) \cite{Kasparian2000,Becker2001,Chen2010a}. From $z = \unit{3.5}{\metre}$, peak intensity and electron density start to rise slowly, eventually reaching maximum values of  $I = \unit{6.2\times10^{13}}{\watt\cdot\centi\metre\rpsquared}$ and $n_{e} = \unit{4\times10^{23}}{\rpcubic\metre}$ at $z = \unit{4.85}{\metre}$. These results correspond respectively to 1.4 and 16 times the clamping values.

The simulated data is thus in excellent agreement with the experiment, both qualitatively and quantitatively, and confirm the existence of a meter-long zone where filaments result in significantly denser plasma channels. Numerical computations also enable us to estimate the peak intensity encountered along $z$, which would otherwise be extremely difficult to measure.

\textit{Discussion}.--- Two main features characterize the studied multifilament bundle: first, we observed a meter-sized region along the beam propagation, shortly before the geometrical focus of the lens, where the plasma density generated by the pulse was significantly higher than the usual clamping value. Secondly, we showed that the same region has a strongly inhomogeneous structure with a few strings, about the size of a standard filament, carrying much of the electron density. These inner structures, which are formed over a meter-long distance, maintain a uniform density and radius over at least several centimeters. We call them \textit{superfilaments} because of the properties they share with standard filaments.

All these features are well reproduced by simulations. The code solves a nonlinear envelope equation, which is a widely used tool to describe known filamentation regimes \cite{Couairon2011}. Numerical computations show that intensity rises moderately above the clamping value for standard filaments in the same area. To date, a similar “over-clamping” behavior has only been observed in very localized regions either in strong focusing conditions \cite{Theberge2006,Kiran2010} or during refocusing cycles of the beam \cite{Gaarde2009,Sun2012}.

The sustaining of filaments with a significant higher plasma density compared to short-scale filaments can be explained by two facts. First, energy losses due to multiphoton absorption are insufficient to balance the energy input due to the focusing action of the lens. As a consequence, intensity and plasma density rise. Second, the defocusing effect of the plasma in the focal region is reduced. Indeed this effect is directly linked to the local electron density gradient. Bringing a large number of filaments very close then smooths the transverse plasma density profile, and therefore results in a weaker defocusing. 

The occurrence of \textit{superfilaments} thus appears strongly correlated to the external focusing action induced by the lens, as evidenced by their sudden disappearance after the geometric focus. However, the moderate beam numerical aperture is also a very important factor leading to this propagation regime. If it is too low, as it is in the case of a collimated beam, then only short-scale filaments are observed, and excess energy is carried by the energy bath \cite{Henin2010}. On the other hand, if it is too high (NA $> 0.1$), a very high intensity can be reached, resulting in the complete ionization of air \cite{Kiran2010a}. However, the length of the dense area is limited in this case to a millimeter-scale.

\textit{Superfilaments} show great promise for the generation of a strong lasing effect from femtosecond filaments, in which the gain medium consists of the filamentation plasma. Up to now, such emission has been reported by several research groups, both in the foward \cite{Liu2013a,Wang2013,Chu2014,Point2014} and backward direction \cite{Luo2003,Kartashov2012b}. Although the underlying mechanisms for population inversion have not been fully elucidated yet, the emission characteristics appear strongly correlated to the value of plasma parameters. The description of a new filamentation regime leading to the formation of dense plasma structures is therefore of the utmost importance. This could pave the way for the development of a strong, coherent, remotely generated ultraviolet light source for long-range probing.

\textit{Conclusion}.--- To conclude, we proved that the global interaction of a dense multifilament bundle leads to the emergence of a new filamentation regime. This regime is characterized by the emergence of a few plasma strings over a meter-long region. These strings have a width comparable to that of short-scale filaments, while being one order of magnitude denser than them on average, and are dubbed \textit{superfilaments}.

\begin{acknowledgments}
\textit{Acknowledgments}.--- The authors would like to thank Dr. Benjamin Forestier for fruitful discussions, and to acknowledge the financial support of the french DGA.
\end{acknowledgments}

\bibliographystyle{apsrev4-1}

\end{document}